\documentclass[iop,apj,twocolumn]{emulateapj}
\usepackage{apjfonts}
\usepackage{graphicx}
\usepackage{epstopdf}

\def\kms{${\rm km\,s}^{-1}$}
\def\cm3{$\rm cm^{-3}$}
\def\Mm1{$\rm Mm^{-1}$}

\shorttitle{TURBULENT CHARACTERISTICS IN A SOLAR QUIESCENT PROMINENCE}
\shortauthors{LEONARDIS ET AL.}
\slugcomment{Submitted 2011 October 13; accepted 2011 November 24; published 2011 January 17}
\journalinfo{The Astrophysical Journal, PAGES, DATE}
\def\andname{}

\begin{document}

\begin{minipage}{2\linewidth}
\vspace{1cm}
\scriptsize \copyright\ 2011. The American Astronomical Society. All rights reserved. Printed in the U.S.A. \normalsize
\end{minipage}

\title{Turbulent characteristics in the intensity fluctuations of a solar quiescent prominence observed by the \textit{Hinode} Solar Optical Telescope }

\author{E.~Leonardis, S.~C.~Chapman, and C.~Foullon}
\affil{Centre for Fusion Space and Astrophysics, Department of Physics, University of Warwick, Coventry, CV4 7AL, United Kingdom}

\begin{abstract}
 We focus on Hinode Solar Optical Telescope (SOT) calcium II H-line observations of a solar quiescent prominence (QP) that exhibits highly variable dynamics suggestive of turbulence. These images capture a sufficient range of scales spatially ($\sim$0.1-100 arc seconds) and temporally ($\sim$16.8 s - 4.5 hrs) to allow the application of statistical methods used to quantify finite range fluid turbulence. We present the first such application of these techniques to the spatial intensity field of a long lived solar prominence. Fully evolved inertial range turbulence in an infinite medium exhibits multifractal \emph{scale invariance} in the statistics of its fluctuations, seen as power law power spectra and as scaling of the higher order moments (structure functions) of fluctuations which have non-Gaussian statistics; fluctuations $\delta I(r,L)=I(r+L)-I(r)$ on length scale $L$ along a given direction in observed spatial field $I$ have moments that scale as $<\delta I(r,L)^p>\sim L^{\zeta(p)}$. For turbulence in a system that is of finite size, or that is not fully developed, one anticipates a generalized scale invariance or extended self-similarity (ESS) $<\delta I(r,L)^p>\sim G(L)^{\zeta(p)}$. For these QP intensity measurements we find scaling in the power spectra and ESS. We find that the fluctuation statistics are non-Gaussian and we use ESS to obtain ratios of the  scaling exponents $\zeta(p)$: these are consistent with a multifractal field and show distinct values for directions longitudinal and transverse to the bulk (driving) flow. Thus, the intensity fluctuations of the QP exhibit statistical properties consistent with an underlying turbulent flow.

\end{abstract}


\section{Introduction}

Solar prominences or filaments in the lower solar corona are relatively cool, dense plasma structures with temperatures of about $10^4$ K. Solar filaments can be seen on the disk, whilst prominences are observed above the solar limb. In practice, they are classified in three main categories according to their location on the Sun, namely active, intermediate  and quiescent. The latter usually occur on the quiet Sun at high latitudes and as a consequence are also known as "polar crown" prominences, while active and intermediate filaments are often observed at low latitudes associated with active regions \citep{Engvold1998}. All prominences originate from filament channels and develop above the polarity inversion line. They show many different morphologies and dynamics \citep[see][for a recent review]{Mackay2010}.

The Hinode Solar Optical Telescope (SOT) provides observations of solar prominences revealing detailed internal dynamics at unprecedented spatio-temporal resolution. In particular, dynamics associated with quiescent prominences (QPs) are seen to exhibit spatio-temporal evolution characterized by high variability \citep{Berger2008}. Most of the QPs in the SOT dataset appear vertically structured and dominated by upward and downward transport of matter; the ascending flows appear dark and are faster (25 \kms on average) than the descending flows (about 10 \kms). The upflows have often been observed to ultimately evolve into vortices \citep{Liggett1984} and are considered to be associated with small scale turbulence \citep{Berger2010}. The prominence is a low-$\beta$ plasma with electron density  $\sim 10^{11}$\cm3 and temperatures up to $\sim 10^4$ K \citep{Tandberg-Hanssen1995}; these typical parameters suggest that the upward flow is supersonic. Indeed, evidence of bow-shock compressions are seen in \citet{Berger2010} and the corresponding Reynolds number is estimated as  $\sim 10^5$. The question then immediately arises as to whether the observed fluctuations do in fact correspond to a turbulent flow.

 Many models have been developed to describe possible scenarios for the production of dynamical structures in the corona. The local magnetic field is suggested to play a key role as it is thought to be the driver of the prominence threads \citep[e.g.,][]{Low1995,Foullon2009,Hershaw2011}. Recently, in strongly inhomogeneous coronal plasma structures, processes such as magneto-thermal convection in solar prominences \citep{Berger2011} and Kelvin-Helmholtz instabilities in the corona \citep{Foullon2011} have been suggested as mechanisms for the generation of dynamical structures. Furthermore, the observed  complexity of the coronal magnetic field may be generated by photospheric turbulence \citep{Abramenko2008,Dimitropoulou2009}. Intriguingly, correlations between outer corona and solar wind have also been found in the statistics of large-scale density fluctuations \citep{Telloni2009a} suggestive that the signature of coronal turbulence is convected with the solar wind plasma \citep{Matthaeus&Goldstein1986}. To distinguish these processes from turbulence evolving in-situ (locally) in the flow, we will apply analysis methods that have been specifically developed to quantify finite range fluid turbulence. 
 
 The characteristic, reproducible properties of a turbulent flow are statistical in nature. They characterize a scale invariance of the statistical properties of fluctuations - that is, these properties are unchanged as we move from scale to scale subject to a rescaling. Thus in a fully evolved magnetohydrodynamic (MHD) turbulent flow in an infinite medium, one finds  power law dependence of the physical observables of the flow - e.g., the velocity and magnetic field fluctuations have power law power spectra over a range of scales, which is identified as the inertial range of the turbulence. As power law power spectra are not unique to turbulence \citep{Sornette} and do not uniquely characterize the scaling of the fluctuations \citep{Chapman2008}, multifractal scaling of the higher order moments (structure functions) of fluctuations is also needed to identify a turbulent flow \citep{Frisch}. These statistical methods have been applied extensively to in-situ observations of the outgoing flow from the solar corona, namely the solar wind, and have established its turbulent character \citep[e.g.,][]{Horbury1997,Sorriso-Valvo1999,Pagel2003,Bruno&Carbone}. However, these in-situ observations are typically single or from a few point in space so that Taylor's hypothesis \citep{Taylor1938} is usually evoked to characterize the scaling properties of the flow. Here, SOT observations of a QP provide a direct observation of the spatial field of fluctuations. 
  
 In this paper we present the first application of these statistical methods to the spatial intensity field of a long lived solar prominence. The SOT images are of intensity, rather than velocity, and so intrinsic to our analysis is the assumption that the moving structures in the images follow the flow, acting as markers or passive scalars for the plasma dynamics (e.g. the intensity measurements can be treated as proportional to the squared density of the plasma flow). This assumption is supported by the correspondence between traceable UV motions and true mass motions, found using combined imaging and Doppler data of prominences \citep{Kucera2003}. Importantly, the QP is of finite physical size, and the turbulence may not be fully developed. Under these circumstances, we anticipate a generalized form of scale invariance, that is, generalized similarity (scale invariance) also known as Extended Self-Similarity (ESS) \citep{Benzi1993}. Generalized similarity has been seen in the fast solar wind \citep{Carbone1996,Hnat2005,Nicol2008,Chapman2009b}, in laboratory simulations of MHD turbulence \citep{Dudson2005,DendyChapman2006} and in hydrodynamics \citep{Grossmann1994,Bershadskii2007}. We find that the intensity fluctuations in the QP do indeed exhibit quantitative features consistent with a finite size turbulent flow, namely, ESS, multifractality and non-Gaussian statistics.
   
\section{The dataset}

The Hinode spacecraft was launched in September 2006 and moves in a sun-synchronous orbit over the day/night terminator, allowing near-continuous observations of the Sun. The SOT on board Hinode is a diffraction-limited Gregorian telescope with a 0.5 m aperture, which is able to provide images of the Sun with an unprecedented resolution up to 0.2 arcsec and cadences between 15 and 30 sec. The Broadband Filter Imager (BFI), one of the four instruments of the Focal Plane Package on the SOT, provides observations over a range of wavelengths (380-670 nm) which distinguish different coronal structures. We use the Ca II H spectral line (396.85 nm) images of a QP observed by the SOT on the north-west solar limb (90W 52N) on November 30th, 2006 (see the time evolution of the QP of interest in animation 1 of \citet{Berger2008}). The time interval considered covers $\sim$4.5 hrs, from 01:00:00 UT to 05:30:00 UT corresponding to about 1000 images with a cadence $\Delta t = 16.8$ sec on average at a spatial resolution of 0.10896 arcsec per pixel, that is, one pixel corresponds to $\Delta r \sim$ 77.22 km on the solar surface; each image is $800 \times 420$ pixels. The images have been calibrated (normalized to the exposure time) and aligned with respect to the solar limb. Furthermore, these specific observations are along a line of sight that is to a good approximation perpendicular to the prominence sheet. 
 
\begin{figure*}
\epsscale{1}
\plotone{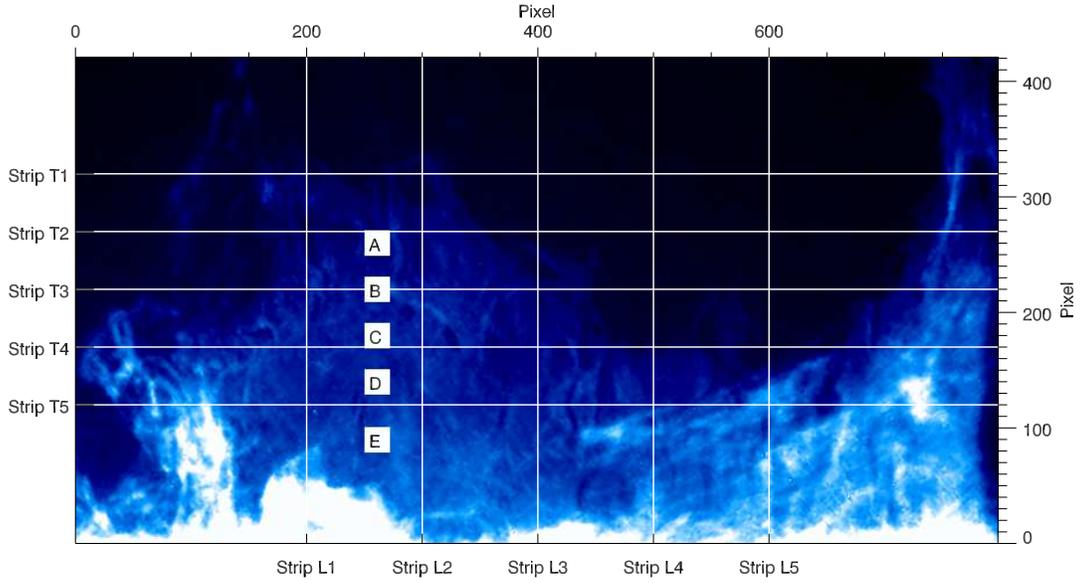}
\caption{QP observed by SOT in the Ca II line on November 30th, 2006 at 01:10:31 UT. Image resolution: 1 pixel $\sim$ 77.22 km on the solar surface. The image has been rotated to the horizontal position with respect to the solar limb. Intensity levels increase from blue to white. The white grid and 5 squares are shown as reference for the analysis in the space domain (transverse strips T1 to T5 and longitudinal strips L1 to L5) and in the time domain (squares A, B, C, D and E).}
\label{figure:fig1}
\end{figure*}

 Figure \ref{figure:fig1} shows the first frame of the dataset. Note the different structures: large scale structures appear brighter at the edge of the prominence while at smaller scales, bright and dark threads alternate within the plasma sheet.
We will examine fluctuations in space by taking differences in intensity along directions longitudinal (vertical) and transverse (horizontal) to the direction of upward/downward flow. This procedure is shown by the  overlaid grid which is made of 10 strips labelled as strips L1 to L5 along the longitudinal direction and strips T1 to T5 along the transverse direction; each strip is 10 pixels wide. We will also examine fluctuations in time, that is, from one image to the next. Five white squares, labelled A to E, with size $21 \times 21$ pixels, indicate the regions over which the respective intensity time series are formed across all the images.

In order to improve statistics we will construct local spatial averages and will present the variation about these averages \citep{Dudok}. The procedure used to analyse the intensity measurements in the strips consists in calculating statistical quantities for small ensembles of 10 neighbouring rows (columns) for each strip along the horizontal (vertical) direction and then performing an average across the strip width. For example, the mean value of the intensities for strip T1 will be the average over the mean values calculated for each of the 10 rows within T1. The same procedure is adopted for the analysis in the time domain: the statistical quantities calculated for each time series associated with the pixels that compose the square are averaged over the 21 x 21 pixels.  

\begin{figure*}
\epsscale{1}
\plotone{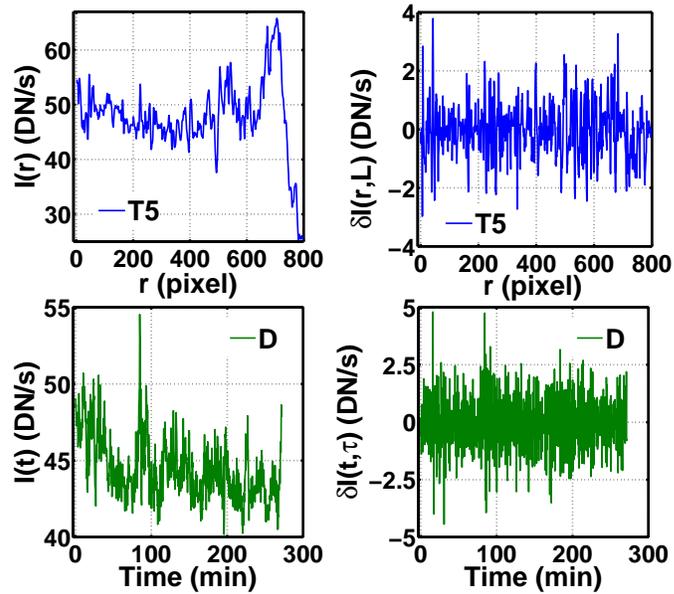}
\caption{\textit{Top panels}: Intensity series $I(r)$ in the space domain for strip T5 (left) and the corresponding first differences $\delta I = I(r+L)-I(r)$ with L=1 pixel (right). \textit{Bottom panels}: Intensity series $I(t)$ in the time domain for square D (left) and the corresponding first differences $\delta I = I(t+\tau)-I(t)$ with $\tau = \Delta t$ = 16.8 sec (right).}
\label{figure:fig2}
\end{figure*}

Figure \ref{figure:fig2} plots the variation in intensity $I(r)$ for strip T5 versus pixel position $r$ (top-left panel) and $I(t)$ for square D versus time (bottom-left panel). They fluctuate strongly in their first differences, which are defined in space as $\delta I(r,L)=I(r+L)-I(r)$ with $L = 1$ pixel (top-right panel) and in time as $\delta I(t,\tau)=I(t+\tau)-I(t)$ with $\tau = \Delta t =$ 16.8 sec (bottom-right panel). This is a typical aspect of a stochastic process including turbulence \citep{Kantz-Schreiber}. 

\section{Power Spectra}
Fully evolved MHD turbulence is self-similar in character exhibiting power law scaling in the power spectrum. We thus analyse the power spectral densities (PSDs) of the intensity measurements both in the time and space domains.
 
\begin{figure*}
\epsscale{1}
\plotone{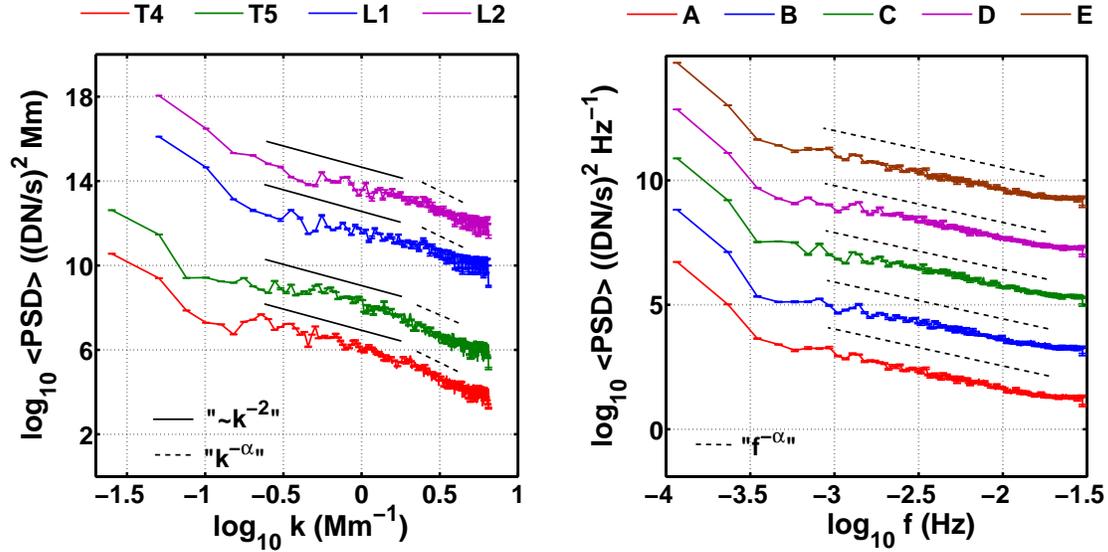}
\caption{Log-log plots of the intensity spectra for strips T4, T5, L1 and L2 (left) and for squares A, B, C, D, and E (right). All the spectra are shifted in the y-direction for clarity. PSDs in the wave number domain (left) reveal two regions with different scaling exponents: $k^{-2}$ (solid line) and $k^{-\alpha}$ (dashed line), while in the frequency domain (right) the PSDs show a single scaling, $f^{-\alpha}$, with spectral index $\alpha$ (dashed line). All the $\alpha$ values are given in Table \ref{table:tab1}.}
\label{figure:fig3}
\end{figure*}

 The left panel in Figure \ref{figure:fig3} shows the PSDs of the intensity measurements for strips T4 and T5 along the transverse direction and L1 and L2 along the longitudinal one. All power spectra are dominated by two main slopes: at small wave numbers the spectra scale as $\sim k^{-2}$ consistent with a Brownian process, that is, additive  noise \citep{Percival-Walden}, while at larger wave numbers the spectra scale as $\sim k^{-\alpha}$ with spectral index $\alpha$ reported in Table \ref{table:tab1} for each strip and suggestive of non-trivial dynamics. The $\alpha$ values are estimated by extracting the gradient of the linear fits to the plots in Figure \ref{figure:fig3} (left) within the wave number ranges 2.43-4.55 \Mm1 for the longitudinal strips and 2.07-4.61 \Mm1 for the transverse strips. The $\alpha$ values found are distinct from -5/3, which is the value expected for the Kolmogorov spectrum for an ideal turbulent flow. This is not surprising since these observations are integrated, line of sight intensity measurements. However, we should still expect line of sight measurements to capture qualitative features of turbulence (such as non-Gaussian fluctuations, multifractal scaling and ESS) whilst not necessary giving the same numerical values of scaling exponents as in-situ point observations.

The time series associated with squares A to E reveal different dynamics: they have power law power spectra in the frequency domain with fitted spectral indices $\alpha$ in the frequency range 1-20 mHz very close to -1 (see Table \ref{table:tab1} and the right panel of Figure \ref{figure:fig3}). This $\sim 1/f$ scaling may be simply attributable to a "random telegraph" process, that is, how a series of uncorrelated pulses or features in the flow moving through the line of sight of the observations \citep{Kaulakys1998,Kaulakys2005}. We estimate the "maximum observable speed" of structures moving past a line of sight as $u = \Delta r / \Delta t \sim \:$ 4.6 \kms. Since the prominence flow has a bulk velocity ($u_{flow} \sim 25$ \kms) larger than $u$ then, at a given pixel, intensity fluctuations are moving too fast for us to observe correlations in time. In other words, the time needed to catch a coherent structure (e.g., up-flows), at fixed space coordinates across two consecutive frames, is much shorter than the cadence, therefore, all the moving flows in the prominence appear decorrelated in time.

\section{Probability distribution}

We now investigate the statistics of the intensity fluctuations in the space domain, $\delta I(r,L)=I(r+L)-I(r)$ with length scale L, and in the time domain, $\delta I(t,\tau)=I(t+\tau)-I(t)$, with time scale $\tau$. Turbulent fluctuations in the inertial range invariably possess a non-Gaussian "heavy tailed" probability density function (PDF) that arises from the intermittent nature of the energy cascade in the flow \citep{Marsch1997,Sorriso-Valvo1999,Hnat2002}.

\begin{figure*}
\epsscale{1}
\center
\plotone{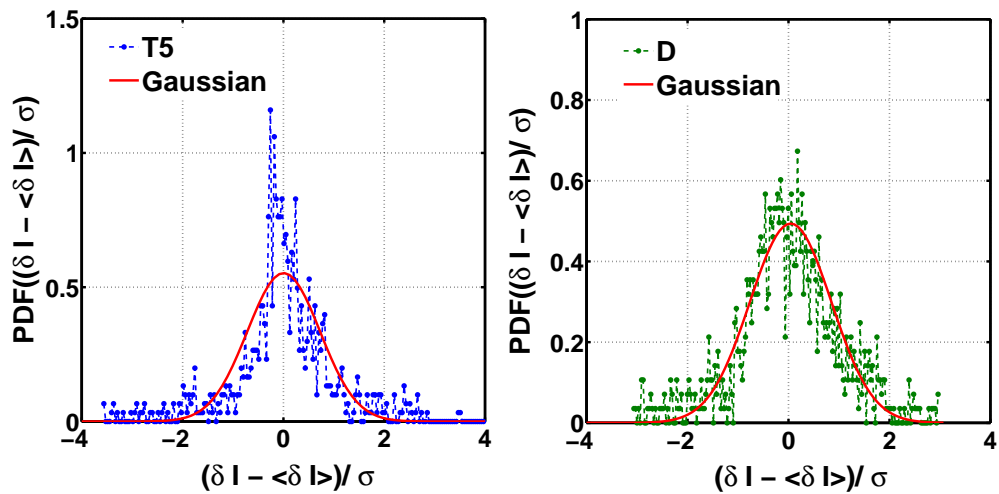}
\caption{PDFs of the intensity fluctuations in space ${\delta}I = I(r+L)-I(r)$ for strip T5 with $L$ = 15 pixel $\sim$ 1.16 Mm (left) and in time ${\delta}I = I(t+\tau)-I(t)$ for square D with $\tau$ = 1.12 min (right). Both PDFs are normalized to the mean <$\delta I$> and the variance $\sigma$ of the intensity. Red solid lines are Gaussian PDFs with $\mu = 0$ and $\sigma = 1$.}
\label{figure:fig4}
\end{figure*}

The left panel in Figure \ref{figure:fig4} shows the PDF of the intensity fluctuations for strip T5, normalized to the mean value $\mu$ and standard deviation $\sigma$, in order to allow comparisons with a Gaussian distribution (solid red line). The PDF of the spatial variations appears to be more peaked compared to the Gaussian distribution. A measure of the "peakedness" of a probability distribution is given by the kurtosis parameter, $k$, defined as $k = <{\delta I >}^4 / \sigma^4$, where $<{\delta I >^4}$ is the fourth moment probability distribution. Since Gaussian distributions have $k=3$, then the excess kurtosis $\overline{k}$ is commonly used, which is defined as $\overline{k} = k-3$.

\begin{figure*}
\epsscale{1}
\center
\plotone{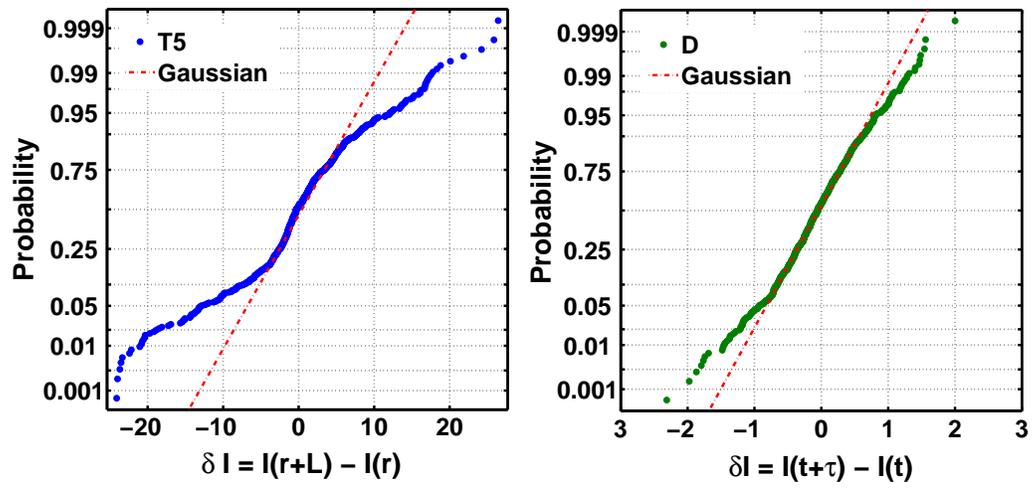}
\caption{Normal probability plots (Q-Q plots against a Gaussian) of the CDFs of the intensity fluctuations for strip T5 in space (left) and square D in time (right). Length and time scales are those in Figure \ref{figure:fig4}. Dashed red lines refer to the probability expected for a Gaussian distribution.}
\label{figure:fig5}
\end{figure*}

The excess kurtosis  $\overline{k}$ calculated for the PDF of strip T5 is 2.44 $\pm$ 0.17 indicating a non-Gaussian distribution. Further evidence of non-Gaussian statistics is given by the normal probability plot of the cumulative distribution function (CDF). This is a quantile-quantile (Q-Q) plot where quantities of the observed CDF (y-axis) are plotted against that of a normal or Gaussian CDF (x-axis). If the data are normal distributed then the normal probability plot of the CDF will be linear, while other distribution types will introduce curvature in the plot. The left panel of Figure \ref{figure:fig5} shows the CDF of strip T5, which does not follow the theoretical function expected for a Gaussian distribution (red dot-dashed line). The temporal fluctuations are of different character: they are more closely described by a Gaussian distribution function. The right panel of Figure \ref{figure:fig4} shows the PDF of the intensity fluctuations for square D which has $\overline{k} = 0.78 \pm 0.16$, indicating statistics very close to Gaussian. Furthermore, the Q-Q plot in the right panel of Figure \ref{figure:fig5} confirms a distribution nearly Gaussian for the temporal fluctuations. The quasi-normal distribution of the fluctuations in the time domain may again be a consequence of the cadence of the observations as discussed in the previous section.

\section{Statistical scaling properties of finite range turbulence}

A key property of turbulence is that it can be characterized and quantified in a robust and reproducible way in terms of the ensemble averaged statistical properties of fluctuations. We can access to the statistical scaling properties of a spatial series, $f(x)$, along a given direction $x$, by constructing differences $\delta f$ with increment $L$:

\begin{equation}
\delta f (x, L) = f(x+L)-f(x) \ 
\label{eq:eqn1}
\end{equation}
on the spatial field. 
Generalized structure functions (GSFs) are a powerful tool to test for statistical scaling and are defined as:
\begin{equation}
S^{p}(L)=\left\langle
\left| \delta f \right|^{p}\right\rangle \
=\int_{-\infty}^{\infty}\left| \delta f \right|^{p}P(\delta f , L) d (\delta f) \ ,
\label{eq:eqn2}
\end{equation}
where the angular brackets indicate an ensemble average over $x$, implying an assumption of approximate statistical homogeneity. Fully developed inertial range turbulence in an infinite medium exhibits the following scaling for the $p$th moment of the GSF:
\begin{equation}
S^p(L) \sim L^{\zeta(p)} \ ,
\label{eq:eqn3}
\end{equation}
where the $\zeta(p)$ are the scaling exponents, which are generally a non linear function of $p$. 

 For the special case of statistical self-similar (fractal) processes one finds a linear form of $\zeta(p)$ in $p$, such that:
\begin{equation}
\zeta(p) = p H \ ,
\label{eq:eqn4}
\end{equation}
where $H$ is the Hurst exponent. 

 In fluid turbulence, we anticipate intermittency, that is $\zeta(p)$ is quadratic in $p$ \citep{Frisch}. Determining the precise $\zeta(p)$ is central to testing turbulence theories. Since we do not have measurements in-situ here, we cannot directly compare our observed $\zeta(p)$ value with predictions of turbulence theories. However we can test whether the $\zeta(p)$ that we observe are non-linear with $p$, consistent with a multifractal, intermittent flow and we discuss this in the next section. 

 First we will focus upon the direct observations of fluctuations in the spatial field as these capture non-trivial correlations in the fluctuations in the flow. The left panels of Figure \ref{figure:fig6} show log-log plots of the averaged 3rd moment of the GSF, $<S_3>$, versus $L_{trans}$ (top panel) for strips T1 to T5 and versus $L_{long}$ (bottom panel) for strips L1 to L6, where  $L_{trans}$ and $L_{long}$ identify the pixel increments of the fluctuations $\delta I(L)$ along the transverse and longitudinal direction respectively. Recall that $<S_3>$ refers to the average over the structure functions calculated for each of the 10 rows (or columns) forming a single strip.
The structure function analysis provides a measurement of the correlation of the fluctuations with length scale L. The increase of the GSFs with L in the left plots of Figure \ref{figure:fig6} thus suggests that the spatial intensity fluctuations of the QP are highly correlated. This is a signature of the presence of coherent structures in the flow.
 In particular, the intensity fluctuations in the longitudinal direction (bottom-left panel) reveal a correlation over a broader range of spatial scales as the coherent structure detected are associated to the up and down flows of the QP, which move along the vertical direction; this is the longitudinal direction in which we expect to see the strongest correlation in a turbulent flow \citep{Frisch}.
Along the transverse direction (top panel of Figure \ref{figure:fig6}), the curves exhibit a knee within a range of length scales of 0.9-2 Mm (dashed black lines). These "break points" delimit the crossover between the small-scale turbulence and the large-scale coherent structures; the above length scales have been attributed to the typical distances between the dark up-flows and seem to be in good agreement with the multi-mode regime of Rayleigh-Taylor instability \citep{Ryutova2010}.

\section{Evidence for ESS and multifractal scaling}

\begin{figure*}
\epsscale{1}
\center
\includegraphics[scale=0.3]{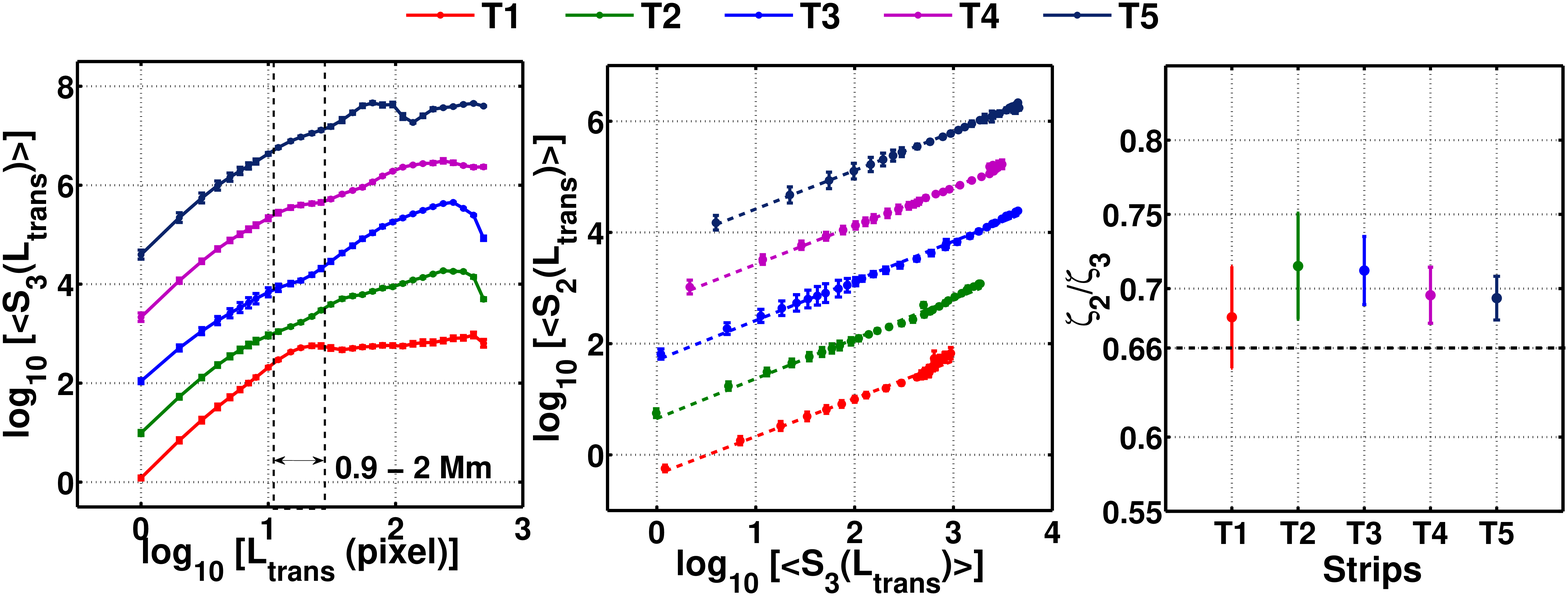}
\includegraphics[scale=0.3]{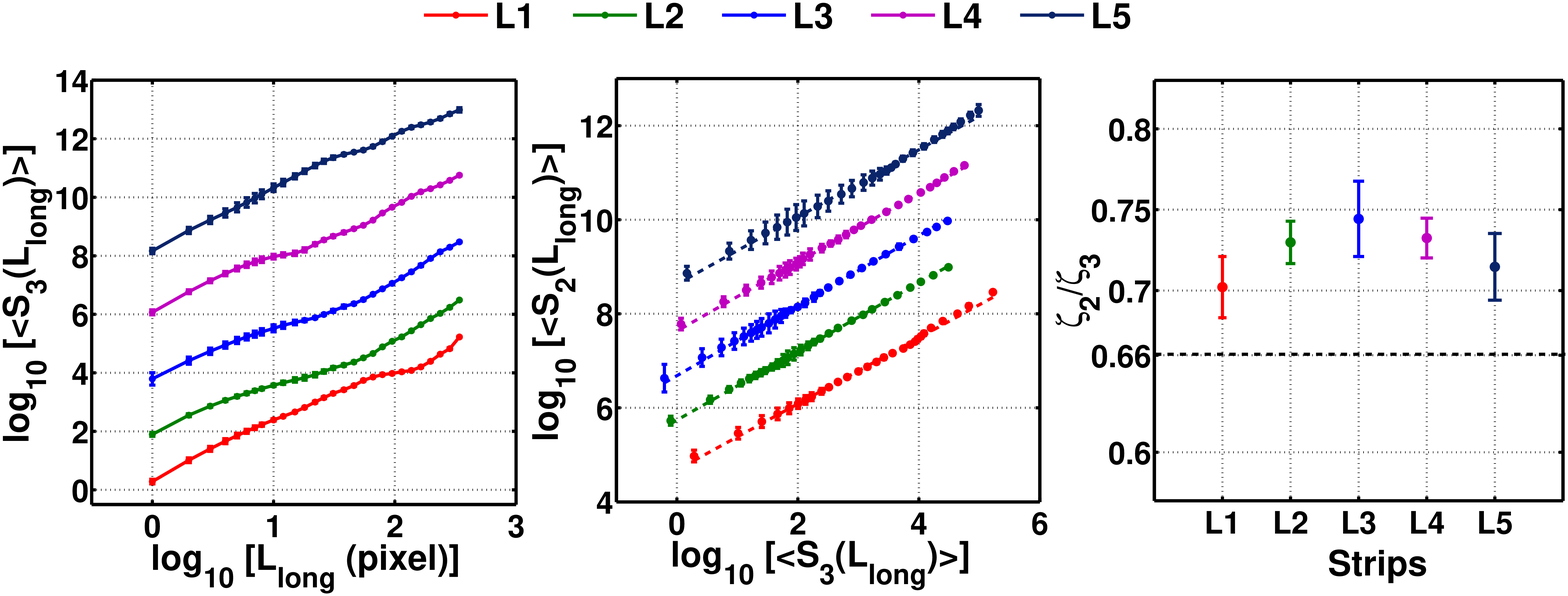}
\caption{\textit{Left panels}: Log-log plot of $<S_3>$ versus $L_{trans}$ for the intensity fluctuations along the transverse direction for strips T1 to T5 (top). Dashed lines delimit the range of scales where the curve exhibit a knee. The bottom panel shows the log-log plot of $<S_3>$ vs. $L_{long}$ for strips L1 to L5 along the longitudinal direction to the flow. All the curves are shifted in the y-direction for clarity. \textit{Middle panels:} Log-log plots of $S_2$ against $S_3$ for all the strips considered showing evidence of ESS. Dashed lines correspond to the linear regression fits across the range $L_{trans}$ = 0.2-5 Mm for the transverse strips (top) and $L_{long}$ = 0.2-10 Mm for strips L1 to L5 (bottom). All the curves are shifted in the y-direction for clarity. \textit{Right panels:} Gradients of the linear fits shown in the respective middle panels, which correspond to the ratio of the scaling exponents ${\zeta(2)}/\zeta(3)$. Notice that the ratios are roughly constant for each direction and different from the value 0.66 for both transverse (top) and longitudinal (bottom) strips. Values of ${\zeta(2)}/\zeta(3) \neq$ 0.66 indicate that ${\zeta(p)}$ is non-linear in $p$ and therefore the prominence flow shows a multifractal character.}
\label{figure:fig6}
\end{figure*}

The GSFs shown in the left panels of Figure \ref{figure:fig6} clearly do not follow the power-law scaling of Equation (\ref{eq:eqn3}). Corrections to this equation indeed have to be taken into account for real turbulent flows for which finite range turbulence effects may arise \citep{Dubrulle2000,Bershadskii2007}; either when turbulence is not completely evolved (low Reynolds number), the data sets are of finite size (realistic cases) or the system is bounded, then symmetries in the flow are broken, and the similarity is lost. Nevertheless, a generalized similarity or ESS has been observed, which suggests a generalized scaling for the $p$th moment of the GSF by replacing $L$ in Equation (\ref{eq:eqn3}) by an initially unknown function $G(L)$, such that

\begin{equation}
S^p(L) \sim G(L)^{\zeta(p)} \ .
\label{eq:eqn7}
\end{equation}
This arises directly from ESS \citep{Benzi1993,Carbone1996}. Comparing structure functions of different order $p$ and $q$, we can then write:

\begin{equation}
S^p(L)=\left[S^q(L)\right]^{\zeta(p)/\zeta(q)}\ .
\label{eq:eqn8}
\end{equation}
A log-log plot of $S^p$ versus $S^q$ will therefore give the ratio of the respective scaling exponents, $\zeta(p)/\zeta(q)$.

The middle panels in Figure \ref{figure:fig6} show the ESS in logarithmic scale of the GSF for $p$=2 and $q$=3. These are straight lines on the log-log of Figure \ref{figure:fig6} thus confirming that Equation (\ref{eq:eqn8}) holds.
The gradient of such plots in the inertial range provides a measurement of the ratio ${\zeta(2)}/{\zeta(3)}$.
Departures of the curves from a linear behaviour occur for length scales outside the inertial range and are associated with large-scale coherent structures in the flow.
Finally, the right panels in Figure \ref{figure:fig6} show ${\zeta(2)}/{\zeta(3)}$ for strips T1 to T5 along the transverse direction (top panel) and for strips L1 to L5 in the longitudinal direction (bottom panel). The error bars provide an estimate of the uncertainty in the gradients of the fitted lines in the inertial range. The ratio ${\zeta(2)}/{\zeta(3)}$ appears to be roughly constant across all the strips and, more interestingly, differs from the value that one would expect if $\zeta(p)$ was linear in $p$, i.e. ${\zeta(2)}/{\zeta(3)} = 2H/3H \sim 0.66$ (see Equation (\ref{eq:eqn4})). The ratios of the scaling exponents found for all the strips are therefore consistent with a non-linear form of the scaling exponent $\zeta(p)$. This is a signature of the multifractal nature of this system which indicates intermittency within the QP flow. 

\begin{figure*}
\epsscale{1}
\center
\plotone{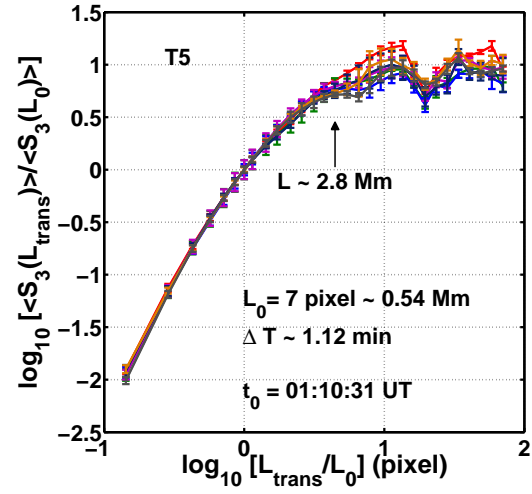}
\caption{Log-log plot of $<{S_3(L_{trans})>/<S_3(L_0)>}$ versus $L_{trans}/L_0$ with $L_0$= 0.54 Mm for strip T5 for seven different frames of the dataset separated by ${\Delta}$T = 1.12 min. $t_0$ is the time of the observation corresponding to the first frame of the dataset. Notice that the various structure functions superimpose up to scale L $\sim$ 2.8 Mm.}
\label{figure:fig7}
\end{figure*}

 The generalized similarity has been tested explicitly in the inertial range of solar wind turbulence by e.g. \citet{Chapman2009a} who formalized Equation (\ref{eq:eqn7}) as follows:
\begin{equation}
S^{p}(L) = [S^p(L_0)] G(L/L_0)^{\zeta(p)}\ ,
\label{eq:eqn9}
\end{equation}
where $L_0$ is some characteristic length-scale of the flow. We finally test this generalized scaling for the intensity fluctuations of strip T5. 
In Figure \ref{figure:fig7} we plot the 3rd moment of the structure function, $S_3$, normalized to a value $L_0 \sim 0.54$ Mm against $L/L_0$ (in logarithmic axes) for 7 consecutive time intervals separated by ${\Delta}T = 1.12$ min and starting at $t_0$= 01:10:31 UT. The choice of the value for the parameter $L_0$ arises from the characteristic width (on average) of the up-flows. The collapse of all the GSFs onto each other within the inertial range indicates the existence of a single scaling function $G(L/L_0)$.
The overlapping of the various GSFs in Figure \ref{figure:fig7} breaks where the effects of large-scale structures become important. This break point for the strip T5 occurs at a length scale of $\sim$2.8 Mm, which corresponds to the width of the large bright structure shown in Figure \ref{figure:fig1} on the right of the prominence and crossed by strip T5 and the length scale of transition to coherent structures in Figure \ref{figure:fig6} (top-left panel).

\section{Conclusions}

We performed the first qualitative test for in-situ turbulence in a QP observed by Hinode/SOT.
We analysed the statistical properties of the spatio-temporal intensity fluctuations associated with the imaged QP from the prospective of a finite sized turbulent system. We found the following:

\
In space:
\

1. The PSDs of the intensity measurements in the space domain exhibit power law scaling suggestive of non-trivial dynamics.

2. The PDFs of the intensity fluctuations are described by non-Gaussian statistics consistent with small-scale MHD turbulence.

3. The GSFs of the intensity fluctuations suggest a generalized scaling for the structure functions with a dependence on a function $G(L)$. They also reveal a high degree of correlation especially along the longitudinal direction to the bulk (driven) flow. Characteristic length scales in the transverse direction have been detected and associated to the characteristic distances between the up-flows.

4. ESS holds for all the strips considered and it is consistent for each direction transverse and longitudinal to the flow as a signature of the generalized similarity expected for finite range turbulent systems.  

5. The ratio of the scaling exponents ${\zeta(2)}/{\zeta(3)}$ is roughly constant for all the strips along each direction and its value is distinct from 0.66, that is the value expected for a fractal system. The prominence flow is therefore multifractal in character, again consistent with in-situ turbulence.

6. The intensity fluctuations in the space domain satisfy the generalized scaling anticipated by ESS and a scaling function $G(L/L_0)$ is observed for different successive time intervals. 

\
In time:
\

7. The PSDs show $\sim 1/f$ scaling, consistent with uncorrelated pulses moving past the line of sight of the observations; The intensity fluctuations are close to Gaussian distributed.

\

The principal aim of this paper has been to explore, for the first time, the possibility of discerning the quantitative signatures of turbulence, namely multifractal or intermittent statistical scaling, within the flows of a long-lived QP.
We have shown how tests for non-Gaussianity, multifractality, scaling and ESS can be applied in order to fully identify and quantify statistical properties of turbulent fluctuations. For these specific intensity measurements we are restricted to a qualitative characterization of the fluctuations since the observations are integrated along the line of sight rather than in-situ in the flow. Despite this constraint, the statistical methods used are powerful tools to test the hypothesis that in-situ flows are turbulent. Their application indeed revealed that the statistical properties of the intensity fluctuations associated with the QP of interest are consistent with a MHD turbulent flow for systems of finite size. This is a clear evidence of in-situ evolving small-scale turbulence within the prominence flow. 

Since many QPs in the Hinode/SOT database exhibit similar dynamics, then this opens up the possibility of using these QPs as a `laboratory for turbulence', to investigate for example finite sized effects on the turbulent flow. The question that immediately arises is whether the flow in these prominences is more generally found to be turbulent. It would be intriguing to determine if or how the presence of turbulence in QPs correlates with their physical properties. Importantly, turbulence is a mechanism by which directed flow is transformed into heat. Heating at the loop footpoints is known to drive condensations at the loop tops \citep[e.g.,][]{Karpen2001}. Rather than heating driven by a coronal or a chromospheric reconnection process, the evidence of turbulence presented here suggests a continuous heating supply that could account for the continuous formation process of QPs.

\acknowledgments
 E. L. and C. F. acknowledge financial support from the UK Engineering and Physical Sciences Research Council and the Science and Technology Facilities Council on the CFSA Rolling Grant, respectively.
 
Hinode is a Japanese mission developed and launched by ISAS/JAXA, with NAOJ as domestic partner and NASA and STFC as international partners. It is operated by these agencies in cooperation with ESA and NSC (Norway).

\bibliography{MyBibliography}

\begin{deluxetable}{cccc}
\tablecaption{Spectral indices}
\tablehead{\colhead{Domain} & \colhead{Data} & \colhead{$\alpha \pm \Delta \alpha$}}
\startdata
 & T4 & 3.17 $\pm$ 0.15 \\
 Wave number & T5 & 2.93 $\pm$ 0.19 \\
 & L1 & 2.73 $\pm$ 0.29 \\
 & L2 & 2.74 $\pm$ 0.37  \\
 \hline 
 & A & 1.21 $\pm$ 0.04 \\
 & B & 1.17 $\pm$ 0.04 \\
 Frequency & C & 1.29 $\pm$ 0.04 \\
 & D & 1.27 $\pm$ 0.04 \\
 & E & 1.20 $\pm$ 0.04 
 
\enddata
\label{table:tab1} 
\end{deluxetable}

\end{document}